\documentclass[12pt,NJP]{iopart}
\usepackage{iopams}
\expandafter\let\csname equation*\endcsname\relax
\expandafter\let\csname endequation*\endcsname\relax
\usepackage{amsmath}
\usepackage{amssymb,amsthm}
\usepackage{amsfonts}
\usepackage{mathtools}
\usepackage{graphicx,subfigure}
\usepackage{epsfig}
\usepackage{multirow}
\usepackage[toc]{appendix}
\usepackage{cite}
\usepackage{color}
\usepackage{srcltx}
\usepackage{soul}
\usepackage{cancel}
\usepackage{subfig}
\usepackage{cite}

\begin{document}
\title{Skewness and Kurtosis in Stochastic Thermodynamics}
\author{Taylor Wampler and Andre C. Barato}
\address{Department of Physics, University of Houston, Houston, Texas 77204, USA}

\begin{abstract}
The  thermodynamic uncertainty relation is a prominent result in stochastic thermodynamics that provides a bound on the fluctuations of any thermodynamic flux, also known as current, in terms 
of the average rate of entropy production. Such fluctuations are quantified by the second moment of the probability distribution of the current. The role of higher order standardized moments such as skewness and kurtosis remains largely unexplored. We analyze the skewness and kurtosis associated with the first passage time of thermodynamic currents within the framework of stochastic thermodynamics. We develop a method to evaluate higher order standardized moments associated with the first passage time of any current. For systems with a unicyclic network of states, we conjecture upper and lower bounds on skewness and kurtosis associated with  entropy production. These bounds depend on the number of states and the thermodynamic force that drives the system out of equilibrium. We show that these bounds for skewness and kurtosis do not hold for multicyclic networks.  We discuss the application of our results to infer an underlying network of states.
\end{abstract}

\maketitle
%%%%%%%%%%%%%%%%%%%%%%%%%%%%%%%%%%%%%%%%%%%%%%%%%%%%%%%%%%%%%%%%%%%%%%%%%%%%%%%%%%%%%%%%%%%%%%%%%%%%%%%%%%%%%%%%%%%%%%%%%%%%%%%%%%%%%%%%%%%%%%%%%%%%%%%%%%%%%%%%%%%%%%%%%%%%%%%%%%%%%%%%%%%%%%%%%%%%%%%%%%
\section{Introduction}
%%%%%%%%%%%%%%%%%%%%%%%%%%%%%%%%%%%%%%%%%%%%%%%%%%%%%%%%%%%%%%%%%%%%%%%%%%%%%%%%%%%%%%%%%%%%%%%%%%%%%%%%%%%%%%%%%%%%%%%%%%%%%%%%%%%%%%%%%%%%%%%%%%%%%%%%%%%%%%%%%%%%%%%%%%%%%%%%%%%%%%%%%%%%%%%%%%%%%%%%%%

The theory of thermodynamics developed in the 19th century is restricted to equilibrium macroscopic systems. Fluctuations are not relevant for such systems and, therefore, are neglected in classical thermodynamics. However, 
many small physical systems with non-negligible fluctuations  such as colloids and single enzymes can be realized in the laboratory today.                  The appropriate 
framework for some of these small nonequilibrium systems is stochastic thermodynamics \cite{seif12}. In this theory, thermodynamic currents such as entropy production, heat, and work are stochastic variables. 

A main goal in stochastic thermodynamics is to find universal relations concerning the  statistics of thermodynamic currents. The most prominent such relation is the fluctuation theorem \cite{seif12} that can be expressed as 
a symmetry for the probability distribution of the entropy production. A more recent development is the thermodynamic uncertainty relation \cite{bara15a}, which is a lower bound on the fluctuations of any thermodynamic current that 
depends only on the average rate of entropy production. Much recent work has been done since the proposal of the thermodynamic uncertainty relation \cite{ging16,piet16,nguy16,piet18,pole16,nyaw16,guio16,piet17b,horo17,pigo17,proe17,maes17,hyeo17,bisk17,bran18,nard18,chiu18,bara18c,dech18,caro19,liu19,guar19,koyu20,ito20,hase21}.

Instead of  stochastic current for a fixed time interval one can consider an observable that is the first passage time to reach a certain threshold current. These observables are equivalent as the 
current distribution contains the information from the first passage time distribution. Both the fluctuation theorem and the thermodynamic uncertainty relation have been expressed in terms of first passage times in \cite{rold15,sait16,neri17} and \cite{ging17a} (see also \cite{garr17}), respectively.

In this paper we analyze skewness and kurtosis associated with the first passage time of a current. We develop a method to calculate the moments associated with the first passage time distribution in terms of 
the transition rates for discrete Markov processes. We  conjecture lower and upper bounds on the skewness and kurtosis associated with the first passage time of entropy for unicyclic networks. For multicyclic networks, these bounds are shown to be violated. We discuss how our bounds are potentially applicable to the problem of inferring a network of states from statistical data of the first passage time of a current. For instance, this problem is relevant in statistical kinetics \cite{moff14}.

The paper is organized as follows. In Sec. \ref{sec2} we define stochastic currents and their first passage times. Sec. \ref{sec3} contains the method we develop to calculate skewness and kurtosis associated with the first passage time distribution. The bounds for unicyclic networks are  discussed in Sec. \ref{sec4}. We show that these bounds are violated  in a multicyclic networks in Sec. \ref{sec5}. We conclude in Sec. \ref{sec6}.

%%%%%%%%%%%%%%%%%%%%%%%%%%%%%%%%%%%%%%%%%%%%%%%%%%%%%%%%%%%%%%%%%%%%%%%%%%%%%%%%%%%%%%%%%%%%%%%%%%%%%%%%%%%%%%%%%%%%%%%%%%%%%%%%%%%%%%%%%%%%%%%%%%%%%%%%%%%%%%%%%%%%%%%%%%%%%%%%%%%%%%%%%%%%%%%%%%%%%%%%%%
\section{Currents and first passage time} 
%%%%%%%%%%%%%%%%%%%%%%%%%%%%%%%%%%%%%%%%%%%%%%%%%%%%%%%%%%%%%%%%%%%%%%%%%%%%%%%%%%%%%%%%%%%%%%%%%%%%%%%%%%%%%%%%%%%%%%%%%%%%%%%%%%%%%%%%%%%%%%%%%%%%%%%%%%%%%%%%%%%%%%%%%%%%%%%%%%%%%%%%%%%%%%%%%%%%%%%%%%
\label{sec2}

\subsection{Stochastic Current} 

Our framework is valid for Markov processes with continuous time and with a finite number of states $\Omega$. The transition rate from a state $i$ to a state $j$ is denoted by $k_{ij}$. In stochastic thermodynamics, we 
typically consider processes such that if $k_{ij}\neq 0$ then $k_{ji}\neq 0$. Furthermore, thermodynamic fluxes, which are the observables of interest in this paper,  are expressed as  stochastic currents. These are 
 functionals of a stochastic trajectory that changes by $\theta_{i,j}$ whenever there is a jump from state $i$ to state $j$. For a stochastic trajectory with fixed time interval $T$, the time-integrated current $\mathcal{J}$
is written  as 
\begin{equation}
\mathcal{J} \equiv \sum_{\textrm{all jumps }l} \theta_{i_l^-,i_l^+},
\label{eqgencurr}
\end{equation}
where $i_l^-$ is the state of the system before a jump $l$ and $i_l^+$ is the state of the system after jump $l$. For currents, the increments   $\theta_{i_l^-,i_l^+}$ are antisymmetric, i.e.,  $\theta_{i_l^-,i_l^+}= - \theta_{i_l^+,i_l^-}$. However, 
the results obtained in Sec. 3 also hold for increments that do not fulfill this property.

In the long time limit $T\to \infty$, the statistics of  $\mathcal{J}$ is described by the scaled cumulant generating function
\begin{equation}
\lambda(z)\equiv \textrm{lim}_{T\to\infty} \langle e^{-z \mathcal{J}}\rangle_T/T,
\end{equation}
where the brackets with subscript $T$ denotes an average over stochastic trajectories with fixed time interval $T$. This scaled cumulant generating function is the maximum eigenvalue of the modified generator \cite{lebo99}
\begin{equation}
\textbf{L}(z)_{ij} \equiv \textrm{e}^{-\theta_{ji}z}k_{ji}(1- \delta_{ij})- \delta_{ij}\sum_{k\neq i} k_{ik}. 
\label{eqgen}
\end{equation}

\subsection{First passage time} 

Instead of the current $\mathcal{J}$ for a trajectory with fixed time $T$ we can consider the first passage  time $\mathcal{T}$ to reach a certain threshold current $J_{t}$.  The statistics of $\mathcal{T}$ in the limit  $J_{t}\to \infty$ is described by another scaled cumulant generating function, given by
\begin{equation}
\rho(s) \equiv \text{lim}_{J_{t} \to \infty} \ln \langle e^{-s \mathcal{T}}\rangle/J_t, 
\end{equation}
where the brackets without any subscript denote an average over stochastic trajectories with fixed threshold current  $J_t$. The derivatives of $\rho(s)$ at $s=0$ give the cumulants associated with the first passage time
$\mathcal{T}$ through the relation 
\begin{equation}
\frac{\kappa_i}{J_t}\equiv(-1)^{i}\frac{d^{i}\rho}{ds^i}(s=0),
\label{eqderi}
\end{equation}
where $\kappa_i$ denotes the cumulant of order $i$ associated with $\mathcal{T}$.

 In particular, we are interested in the third standardized moment skewness 
\begin{equation}
S \equiv |J_t|^{1/2}\frac{\kappa_3}{\kappa_2^{3/2}}
\label{eqdefS}
\end{equation}
and the fourth standardized moment (excess) kurtosis 
\begin{equation}
K \equiv  |J_t|\frac{\kappa_4}{\kappa_2^{2}}.
\label{eqdefK}
\end{equation}
There are two relevant comments about these definitions. First, skewness and kurtosis are typically defined without the $J_t$ factors. We have scaled $S$ and $K$ by factors of $J_t$ so that they are finite in the limit $J_t\to\infty$, in 
agreement with Eq. \eqref{eqderi}. Hence, these quantities could be called scaled skewness and scaled kurtosis. Second, we have used the absolute value of $J_t$ in our definitions for the following reason.  There are two possible 
first passage time distributions. One is the distribution for typical events that corresponds to a threshold current $J_t$ that has the same sign as the sign of the average of the stochastic current  $\langle \mathcal{J}\rangle_T$. The other is the first passage time distribution for 
rare events that corresponds to a $J_t$ that has an opposite sign in relation to the sign of $\langle \mathcal{J}\rangle_T$. Here we consider the first passage distribution for typical events with $J_t$ and $\langle \mathcal{J}\rangle_T$ with the same sign.  Since we do not know  the sign of the average current $\langle \mathcal{J}\rangle_T$ in terms of the transition rates, our method must  work for both cases, positive and negative signs.

In order to account for the sign of $J_t$ we consider the the first derivative in Eq. \eqref{eqderi},  
\begin{equation}
\textrm{sign}J_t=-\textrm{sign}(\rho')\equiv \epsilon,
\end{equation}
where the prime denotes a derivative at $s=0$. We can now write $S$ and $K$ in terms of derivatives of $\rho$ in a way consistent with both cases, positive and negative $J_t$.
Skewness in Eq. \eqref{eqdefS} and kurtosis in Eq. \eqref{eqdefK} can be written in terms of derivatives of $\rho(s)$ at $s=0$ from Eq. \eqref{eqderi}, which gives the following relations,
\begin{equation}
S= -\epsilon\rho'''/(\epsilon\rho'')^{3/2}
\label{eqS}
\end{equation}
and 
\begin{equation}
K= \epsilon\rho''''/(\epsilon\rho'')^{2},
\label{eqK}
\end{equation}
where the primes denote derivatives at $s=0$. We reiterate that with $\epsilon$ we always consider the first passage time for typical events, with $J_t$ and the average current $\langle\mathcal{J} \rangle_T$ with the same sign, 
independent of whether this sign is positive or negative.

The statistics of a stochastic current $\mathcal{J}$ for fixed time $T$ are connected to the statistics of the first passage time $\mathcal{T}$ for fixed threshold current $J_t$.
This connection is represented by the following relation between $\lambda(z)$ and $\rho(s)$. Consider the characteristic polynomial associated with the modified generator in Eq. \eqref{eqgen} 
\begin{equation}
\Xi(s, z)\equiv det(s\mathbb{I} - \textbf{L}(z)),
\label{Eqchar}
\end{equation}
where $\mathbb{I}$ is the identity matrix and $\textbf{L}(z)$ is the matrix defined in Eq. \eqref{eqgen}. The maximal root of this polynomial in $s$ is the scaled cumulant generating function 
for a current $\lambda(z)$. The scaled cumulant generating function  for the first passage time $\rho(s)$ fulfills the relation \cite{ging17a}  
\begin{equation}
\Xi(s, \rho(s))=0.
\label{Eqrho1}
\end{equation}
In principle, we can use this equation to determine $\rho(s)$ in terms of the transition rates $k_{ij}$. From $\rho(s)$ we can determine skewness $S$ from Eq. \eqref{eqS} and kurtosis $K$ from Eq. \eqref{eqK}.
However, the explicit form of $\rho(s)$ in terms of the transition rates $k_{ij}$ is hard to obtain even for systems with a small number of states. In the next section, we introduce a method to circumvent this issue, i.e.,
one can determine $S$ and $K$ in terms of the transition rates without the explicit form of $\rho(s)$.

%%%%%%%%%%%%%%%%%%%%%%%%%%%%%%%%%%%%%%%%%%%%%%%%%%%%%%%%%%%%%%%%%%%%%%%%%%%%%%%%%%%%%%%%%%%%%%%%%%%%%%%%%%%%%%%%%%%%%%%%%%%%%%%%%%%%%%%%%%%%%%%%%%%%%%%%%%%%%%%%%%%%%%%%%%%%%%%%%%%%%%%%%%%%%%%%%%%%%%%%%%
\section{Method to calculate standardized moments}
%%%%%%%%%%%%%%%%%%%%%%%%%%%%%%%%%%%%%%%%%%%%%%%%%%%%%%%%%%%%%%%%%%%%%%%%%%%%%%%%%%%%%%%%%%%%%%%%%%%%%%%%%%%%%%%%%%%%%%%%%%%%%%%%%%%%%%%%%%%%%%%%%%%%%%%%%%%%%%%%%%%%%%%%%%%%%%%%%%%%%%%%%%%%%%%%%%%%%%%%%%
\label{sec3}

Our method follows the same rationale of a the method by Koza \cite{koza99} to obtain the cumulants associated with the stochastic current for a fixed time without the explicit calculation of $\lambda(z)$. In Koza's method
one can obtain the derivatives of $\lambda(z)$ at $z=0$ in terms of the coefficients of the characteristic polynomial in Eq. \eqref{Eqchar}: there is no need to find the root of the polynomial $\lambda(z)$, which is a much more
complicated problem. These coefficients in Koza's method $B_i(z)$ are defined through the relation $\Xi(s, z)= \sum_{i=0}^\Omega B_i(z)s^i$.

Consider the characteristic polynomial $\Xi(s, \rho)$, defined in Eq. \eqref{Eqchar} as a function of two independent variables $s$ and $\rho$. A Taylor expansion of Eq. \eqref{Eqrho1}  in $\rho$ leads to 
\begin{equation}
\sum_{i=0}^{\infty} {c_i(s)\rho(s)^i}=0,
\label{Eqrhoc}
\end{equation}
where $c_i$ is the $i$ derivative of $\Xi(s, \rho)$ with respect to $\rho$.  We note that these derivatives are taken with $\rho$ as an independent variable. However, we now consider $c_i(s)$ and $\rho(s)$ as functions of $s$ in Eq. \eqref{Eqrhoc}.
For our final derivation we use the relation \cite{ging17a}
\begin{equation}
\rho(s=0)=0,
\end{equation}
which is true for the typical first passage time distribution that corresponds to a threshold current with the same sign as the average current. Using this relation and taking derivatives with respect to $s$ at $s=0$ 
in Eq. \eqref{Eqrhoc} we obtain, 
\begin{align}
 & \rho^{'} = -\frac{c_0^{'}}{c_1} \nonumber\\
 & \rho^{''} = -\frac{c_0^{''}+2 c_2 (\rho^{'})^2+2 c_1^{'} \rho^{'}}{c_1} \nonumber\\
  & \rho^{'''} = -\frac{c_0^{(3)}+ 6 c_2 \rho^{'} \rho^{''} +3 c_1^{'} \rho^{''} +3 \rho^{'} c_1^{''} +6 c_3 (\rho^{'})^3 +6 c_2^{'} (\rho^{'})^2}{c_1} \nonumber\\
  & \rho^{''''} = -\frac{c_0^{(4)}+ 8 c_2 \rho^{'} \rho^{(3)} +4 c_1^{'} \rho^{(3)} +4 \rho^{'} c_1^{(3)} +6 c_2 (\rho^{''})^2 +6 c_1^{''} \rho^{''}}{c_1}\nonumber\\
  & -\frac{ 36 c_3 (\rho^{'})^2 \rho^{''} + 24 {c_2}^{'} \rho^{'} \rho^{''} +12 (\rho^{'})^2 {c_2}^{''} + 24 {c_4} (\rho^{'})^4 + 24 {c_3}^{'} (\rho^{'})^3}{c_1}.
  \label{Eqderi}
\end{align}
This equation together with Eq. \eqref{eqS} and Eq. \eqref{eqK} allow us to obtain skewness $S$ and kurtosis $K$ in terms of the coefficients $c_i(s)$ in Eq. \eqref{Eqrhoc}. 

Summarizing, one can calculate $S$ and $K$ in terms of the transition rates $k_{ij}$ with the following algorithm. First, evaluate the characteristic polynomial in Eq. \eqref{Eqchar}. Second, obtain the coefficients $c_i(s)$ up to $i=4$ 
in the  Taylor expansion in Eq. \eqref{Eqrhoc}. Third, calculate the derivatives of $\rho$ at $s=0$ in \eqref{Eqderi}. The skewness $S$ can be obtained with Eq. \eqref{eqS} and the kurtosis $K$ can be obtained with Eq. \eqref{eqK}.
Hence, one can obtain the skewness and the kurtosis associated with first passage time of a current in terms of the transition rates without explicitly calculating roots of the polynomial in Eq. \eqref{Eqchar}. This method is our first main result.
The method can also be used to calculate higher order cumulants by simply calculating higher order derivatives of $\rho$. We reiterate that the method is not restricted to currents that have antisymmetric increments but also applies to any 
observable  of the form given in Eq. \eqref{eqgencurr}.

%%%%%%%%%%%%%%%%%%%%%%%%%%%%%%%%%%%%%%%%%%%%%%%%%%%%%%%%%%%%%%%%%%%%%%%%%%%%%%%%%%%%%%%%%%%%%%%%%%%%%%%%%%%%%%%%%%%%%%%%%%%%%%%%%%%%%%%%%%%%%%%%%%%%%%%%%%%%%%%%%%%%%%%%%%%%%%%%%%%%%%%%%%%%%%%%%%%%%%%%%%
\section{Bound for unicyclic networks}
%%%%%%%%%%%%%%%%%%%%%%%%%%%%%%%%%%%%%%%%%%%%%%%%%%%%%%%%%%%%%%%%%%%%%%%%%%%%%%%%%%%%%%%%%%%%%%%%%%%%%%%%%%%%%%%%%%%%%%%%%%%%%%%%%%%%%%%%%%%%%%%%%%%%%%%%%%%%%%%%%%%%%%%%%%%%%%%%%%%%%%%%%%%%%%%%%%%%%%%%%%
\label{sec4}

\subsection{Unicyclic networks and bound on second cumulant}

We now consider  a unicyclic network with $\Omega$ states. The transition rate from state $i$ to state $i+1$ is denoted $k_i$ and  the transition rate from state $i$ to state $i-1$ is denoted $\overline{k}_i$. 
The unicyclic  network has periodic boundary conditions, the transition rate from $i=\Omega$ ($i=1$) to $j=1$ ($j=\Omega$) is denoted $k_\Omega$ ($\overline{k}_1$). Possible physical interpretations for such a model 
are an enzyme with a single cycle or a colloid on a ring.  The thermodynamic affinity is defined as 
\begin{equation}
A\equiv \prod_{i=1}^{\Omega}( k_i/\overline{k}_i).
\label{eqaff}
\end{equation}
 If $A=0$ the system is in equilibrium and if $A\neq 0$ the system is out of equilibrium. For an enzyme that burns one ATP in a cycle, the thermodynamic affinity is the free energy of ATP hydrolisis and for a colloidal particle on a ring it is the work done by the force 
that drives the particle in one loop, both in units of $k_BT$, where $k_B$ is Boltzmann's constant and $T$ is the temperature. Without loss of generality we will consider the case $A\ge 0$. 
 
 The entropy production has the following increments, $\theta_{\Omega,1}=-\theta_{1,\Omega}=A$. For any other jumps the increment is zero. We consider the entropy since this is a general current that can also be analyzed in multicyclic networks. However, since the network of states is unicyclic there is only one independent current due to Kirchhoff's law and different increments lead to the same results up to a rescaling factor. We also consider the cases $\Omega=1$ and $\Omega=2$. For $\Omega=1$ we simply have a biased random walk that jumps to the right with rate $k\textrm{e}^{A}$ and to the left with rate $k$. For $\Omega= 2$ there must be two links between the two states. The modified generator in this case is a $2\times2$ matrix with elements $L_{11}= -(k_1^++k_1^-)$, $L_{22}= -(k_2^++k_2^-)$, $L_{21}= k_1^++k_1^-\textrm{e}^{Az}$, and 
 $L_{12}= k_2^+\textrm{e}^{-Az}+k_2^-$.

 Before we present our results we mention the following existing bound. Consider the following quantity related to the second cumulant of the first passage time distribution, 
 \begin{equation}
R\equiv  J_t\frac{\langle(\mathcal{T}-\langle\mathcal{T}\rangle)^2\rangle}{\langle\mathcal{T}\rangle^2},
\end{equation}
 which is known as randomness parameter in statistical kinetics \cite{moff14}. For a fixed affinity $A$ and  number of states $\Omega$, there is a lower bound on $R$ for unicyclic networks \cite{bara15},
\begin{equation}
R\ge  A\Omega^{-1}\coth[A/(2\Omega)]. 
\label{eqboundR}
\end{equation}
The randomness parameter is a function of $2\Omega$ transition rates. If we use the affinity $A$ to fix one of the rates, it becomes a function of $2\Omega-1$ variables. The minimum of $R$ for fixed $A$ and $\Omega$, expressed on the right hand side of Eq. \eqref{eqboundR}, is achieved for uniform rates, i.e., $k_i= k \textrm{e}^{A/\Omega}$ and $\overline{k}_i= k $ for all $i$. Hence, $R$ is minimized for an asymmetric random walk (ARW).

\subsection{Bounds on  Skewness}

We used the method introduced in Sec. \ref{sec3} to evaluate the skewness and kurtosis for unicyclic networks as functions of the transition rates. Let us first consider the skewness $S$. If we consider an ARW, which corresponds to uniform rates, with affinity $A$ and number of states $\Omega$ the skewness is given by
\begin{equation}
S_{ARW}(A, \Omega) = \frac{2\sqrt{A}}{\sqrt{ \Omega}}\frac{1+4e^{A/ \Omega}+e^{2A/ \Omega}}{\sqrt{(-1+e^{A/ \Omega})(1+e^{A/ \Omega})^3}}.
\label{eqskewarw}
\end{equation}

\begin{figure}
\subfigure[$\Omega= 3$]{\includegraphics[width=75mm]{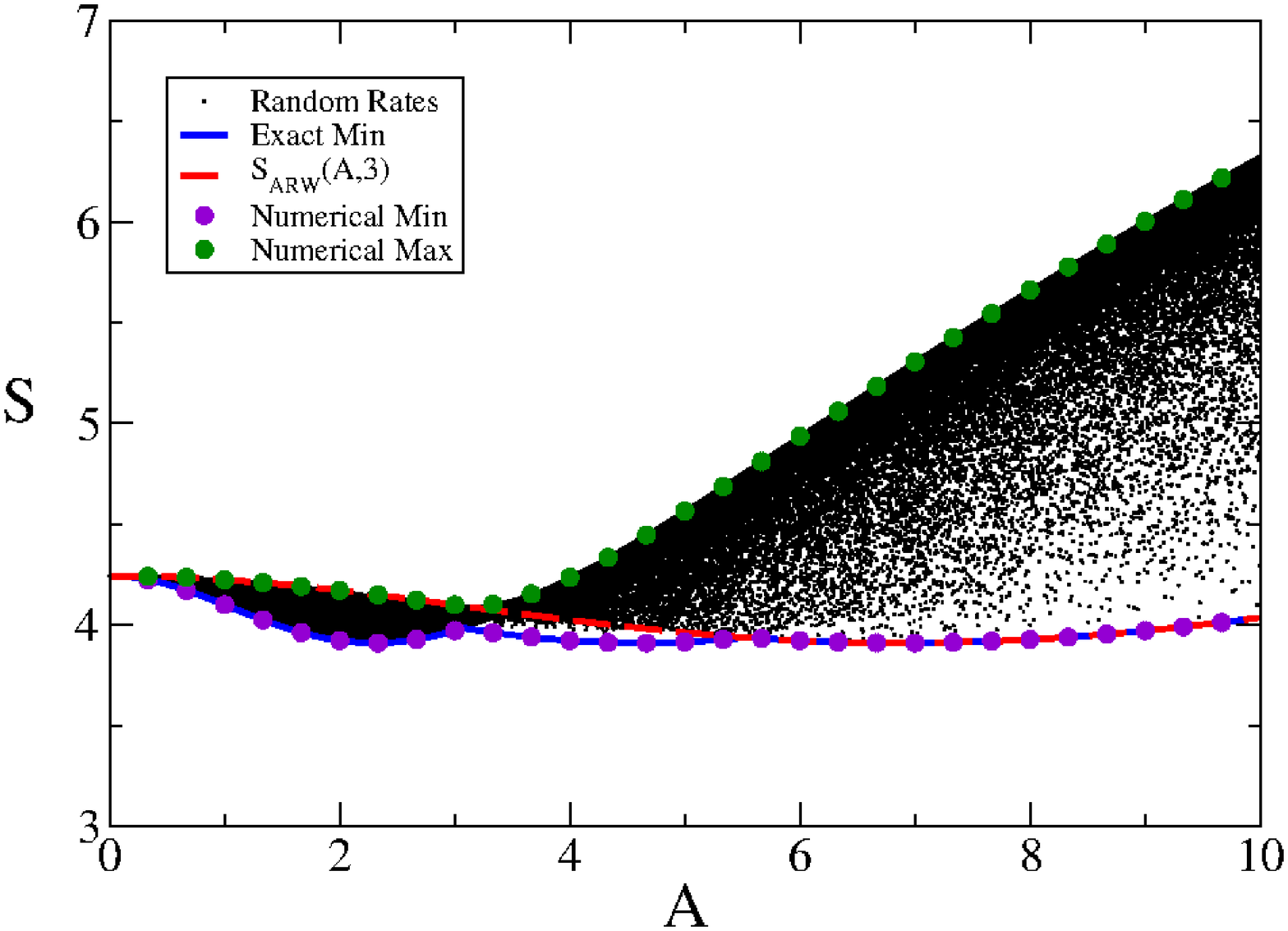}}
\subfigure[$\Omega= 4$]{\includegraphics[width=75mm]{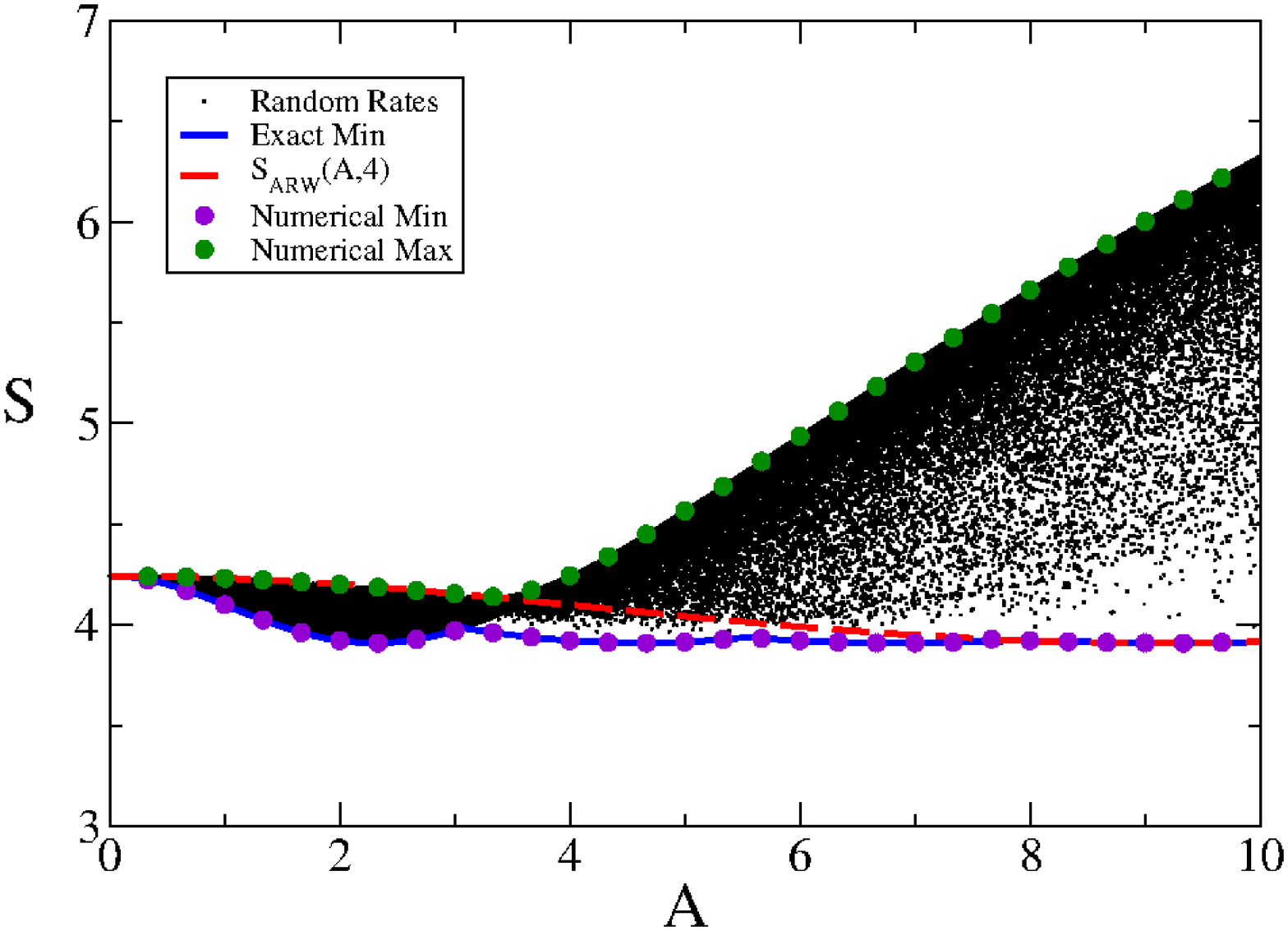}}
\vspace{-3mm}
\caption{Bounds on skewness for (a) $\Omega=3$ and  (b) $\Omega=4$. We have plotted $S$ for randomly chosen rates, the exact lower bound in Eq. \eqref{eqsmin}, $S_{ARW}(A,\Omega)$, the numerical minimum of $S$, and the numerical maximum of $S$.}

\label{fig1}
\end{figure}

An important difference between the expression for  $S$ in Eq. \eqref{eqskewarw} and the lower bound on $R$ in Eq. \eqref{eqboundR}  is that the lower bound on $R$ is a decreasing function of $\Omega$  while $S_{ARW}(A, \Omega)$ is not a decreasing function of $\Omega$. For a given value of the affinity $A$ there is an optimal value of the integer $\Omega$ that minimizes $S_{ARW}(A, \Omega)$. For  $0\le A\le A_1^S$ the function  $S_{ARW}(A, 1)$ is the minimal one,
where $A_1^S$ is the solution of the transcendental equation $S_{ARW}(A_1^S, 1)=S_{ARW}(A_1^S, 2)$. In general,  $S_{ARW}(A, i)$ is minimal for $A_{i-1}^S\le A\le A_i^S$, where $A_i^S$ is the solution of the transcendental equation $S_{ARW}(A_i^S, i)=S_{ARW}(A_i^S, i+1)$. Therefore, for a system with $\Omega$ states $S_{ARW}(A, \Omega)$ cannot be the lower bound for all $A$, since it is also possible to choose rates such that a system with $\Omega$ states behaves like an ARW with less than $\Omega$ states.

\begin{figure}
\subfigure[Lower bound for $\Omega= 4$]{\includegraphics[width=75mm]{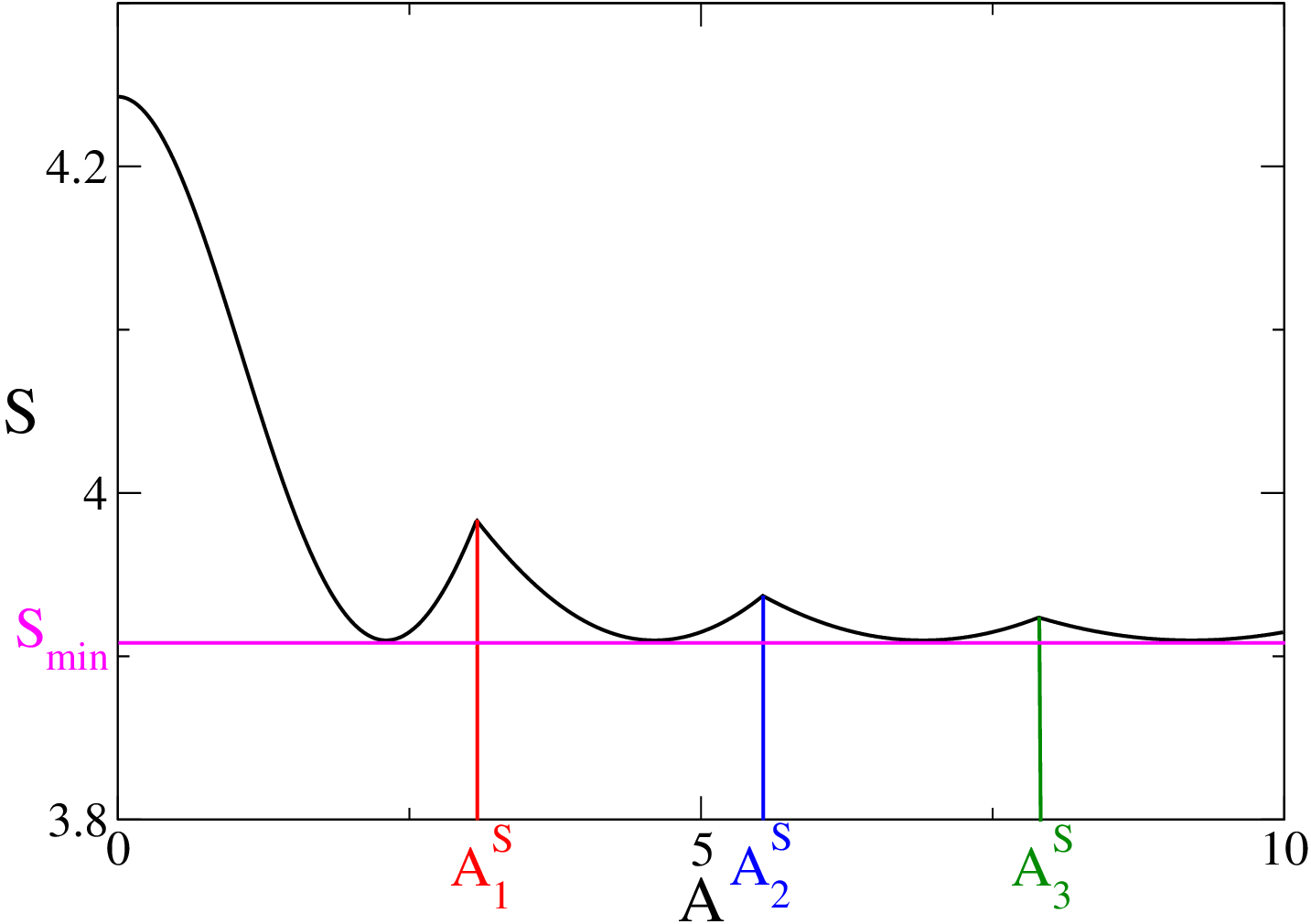}}
\subfigure[Upper bound]{\includegraphics[width=75mm]{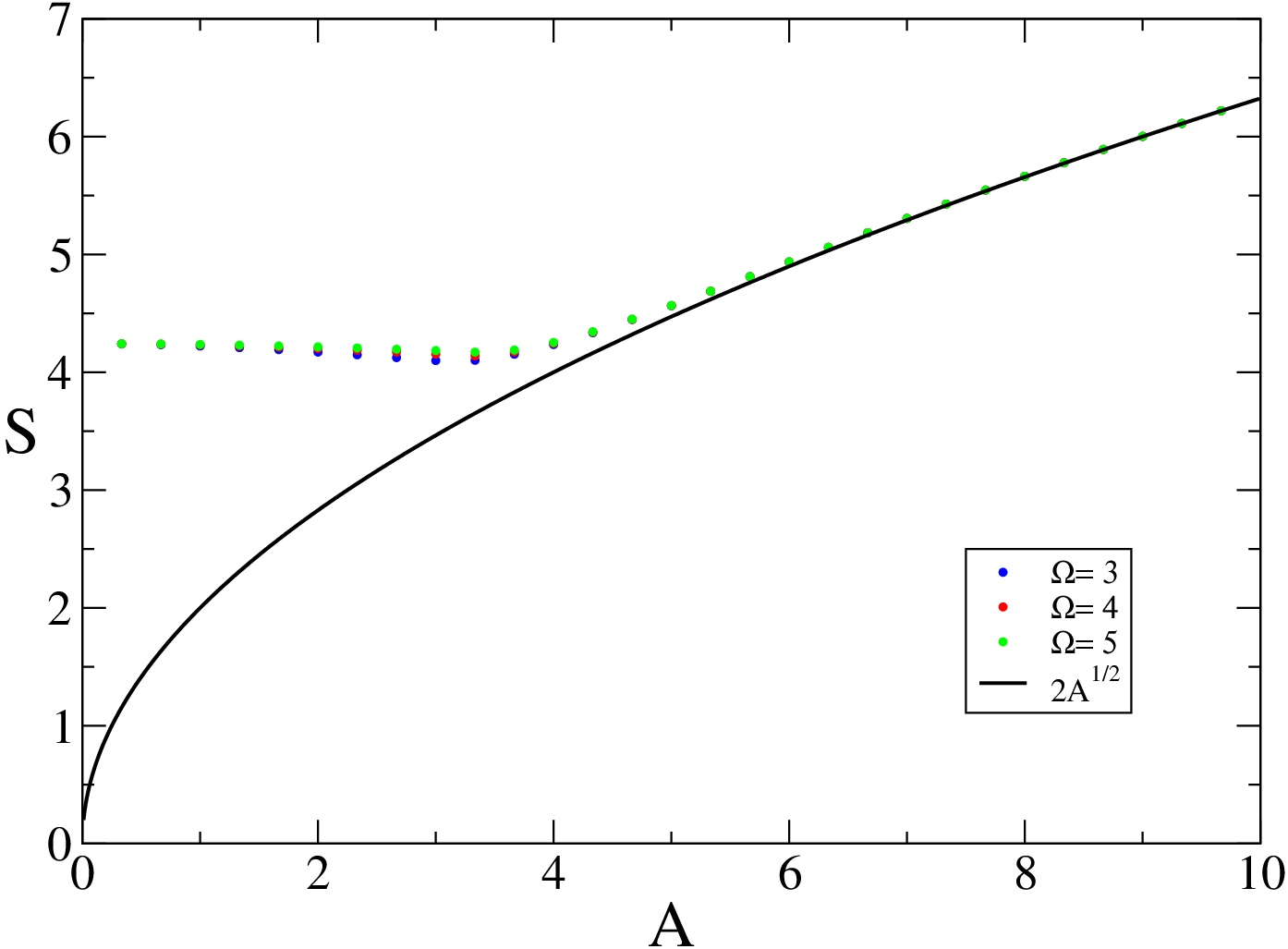}}
\vspace{-3mm}
\caption{(a) llustration of the kinks on the lower bound on $S$ for $\Omega=4$. (b) Illustration of the kink and the asymptotic behavior for large $A$ of the upper bound on $S$.}
\label{fig2}
\end{figure}

Preforming a numerical investigation up to $\Omega=8$ we arrive at the conjecture of the following lower bound on the skewness for fixed affinity $A$ and number of states $\Omega$,
\begin{equation}
S \le \textrm{min}_{N\le\Omega}\{S_{ARW}(A, N)\}.
\label{eqsmin}
\end{equation}
This bound and our numerical investigation is illustrated in Fig. \ref{fig1} for $\Omega=3$ and $\Omega=4$. The lower bound for a system with $\Omega$ states has $\Omega-1$ kinks, 
which happen at points $A_i^{S}$, with $i=1,\ldots,\Omega-1$. For $A>A_{\Omega-1}^{S}$ the lower bound becomes  $S_{ARW}(A, \Omega)$. These kinks are illustrated in Fig. \ref{fig2}(a) for 
the case $\Omega=4$. In our numerical investigation we have performed numerical minimization of $S$ 
with fixed affinity $A$ and number of states $\Omega$. We have also evaluated $S$ for random values for the transition rates and they all stay above the lower bound.

The skewness $S$ has an absolute minimum $S_{min}$, which is independent of the affinity $A$ and the number of states $\Omega$. Interestingly,  for a system with $\Omega$ states, this minimum is reached in each of the $\Omega$ pieces of the lower bound  in Eq. \eqref{eqskewarw}, as shown for $\Omega=4$ in Fig. \ref{fig2}(a). This minimum value is $S_{min}\approx 3.90973$.

Besides this lower bound it turns out that the skewness also has an upper bound illustrated in Fig. \ref{fig2}(b). This upper bound has one kink at $A=A^S_*$ that depends on system size $\Omega$. This kink happens at a different point than the points $A_i^S$ for the kinks of the lower bound. We have determined numerically the value of $A^S_*$ for the upper bound up to $\Omega=8$, as shown in table \ref{tab1}.  Interestingly, an ARW can be both an upper and a lower bound on the skewness depending on the value of the affinity $A$. In particular, $0\le A\le A^S_*$ it is an upper bound and for $A\ge A^S_{\Omega-1}$ it is a lower bound. 

For  $A\ge A^S_*$ we can only determine the upper bound numerically and we do not know its analytical form.  The asymptotic form of the upper bound for large affinity is $2\sqrt{A}$ as illustrated in Fig. \ref{fig2}(b). This asymptotic form corresponds to an ARW with $N=1$ in the large $A$ limit, i.e., $S_{ARW}(A,1)$ defined in Eq. \eqref{eqskewarw} behaves as $2\sqrt{A}$  for large $A$.

\begin{table}
\begin{center}
\begin{tabular}{ |c|c|c| } 

 \hline
$\Omega$ & $A^S_*$ & $A^K_*$ \\ 
 \hline
 2 & 2.84 & 3.27 \\ 
 \hline
 3 & 3.15 & 3.64 \\ 
 \hline
 4 & 3.32 & 3.84 \\
 \hline
 5 & 3.41 & 3.95 \\
 \hline
 6 & 3.47 & 4.02 \\
 \hline
 7 & 3.51 & 4.07 \\
 \hline
 8 & 3.54 & 4.10 \\
 \hline
\end{tabular}
\end{center}
\caption{Points for the kinks of the upper bounds for different system sizes.}
\label{tab1}
\end{table}

In the linear response regime close to equilibrium the lower and upper bounds tend to the same value, as shown in Fig. \ref{fig1}. In this regime the skewness has a fixed value given by $S= 6/\sqrt{2}$. This value can be obtained by taking the limit $A\to 0$ of
 $S_{ARW}(A, \Omega)$ in Eq. \eqref{eqskewarw}.

 \subsection{Bounds on  kurtosis}

\begin{figure}
\subfigure[$\Omega= 3$]{\includegraphics[width=75mm]{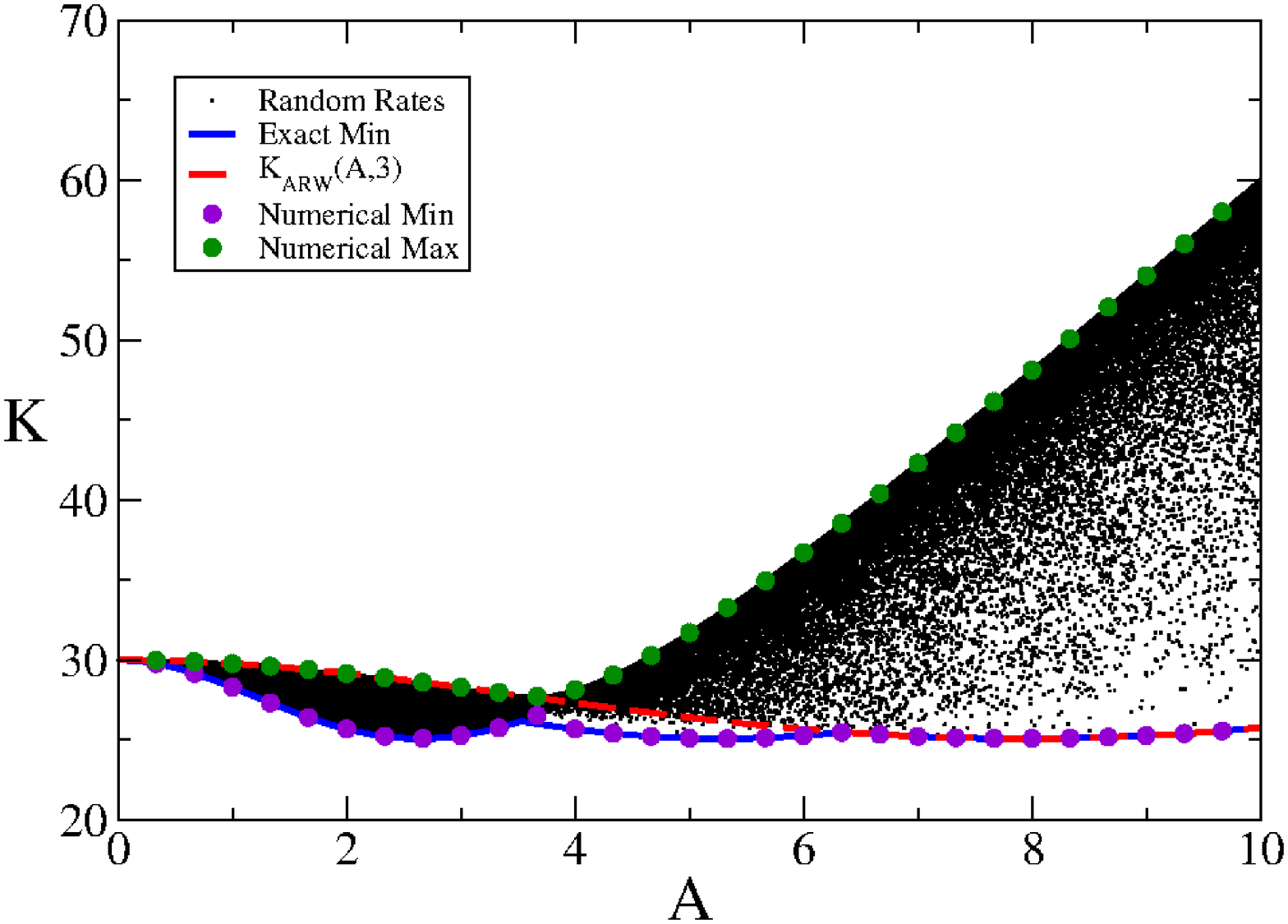}}
\subfigure[$\Omega= 4$]{\includegraphics[width=75mm]{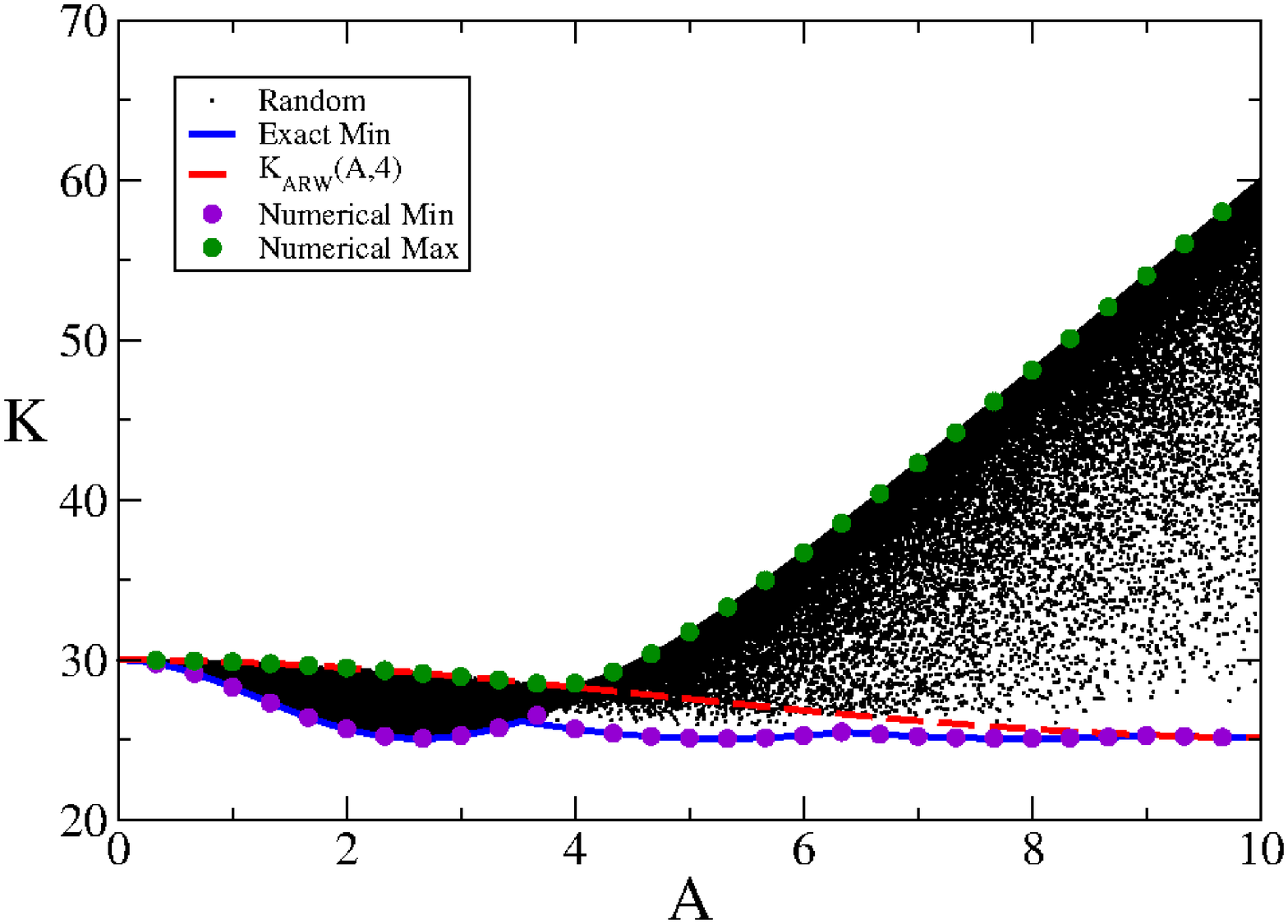}}
\vspace{-3mm}
\caption{Bounds on kurtosis for $\Omega=3$ and $\Omega=4$. We have plotted $K$ for randomly chosen rates, the exact lower bound in Eq. \eqref{eqkmin}, $K_{ARW}(A,\Omega)$, the numerical minimum of $K$, and the numerical maximum of $K$.}
\label{fig3}
\end{figure}

\begin{figure}
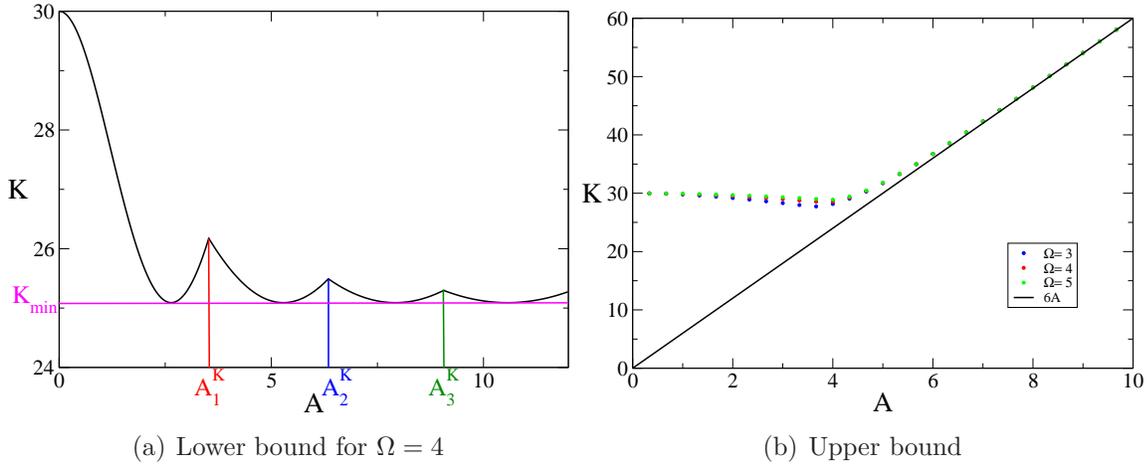

\subfigure[Lower bound for $\Omega= 4$]{\includegraphics[width=75mm]{Fig4a.eps}}
\subfigure[Upper bound]{\includegraphics[width=75mm]{Fig4b.eps}}
\vspace{-3mm}
\caption{(a) Illustration of the kinks of the lower bound on $K$ for $\Omega=4$. (b) Illustration of the kink and the asymptotic behavior for large $A$ of the upper bound on $K$.}
\label{fig4}
\end{figure}

Similar bounds also hold for the kurtosis $K$, as show in Fig. \ref{fig3}. There is a lower bound for fixed $A$ and $\Omega$ given by 
\begin{equation}
K\le \textrm{min}_{N\le\Omega}\{K_{ARW}(A, N)\}.
\label{eqkmin}
\end{equation}
where,
\begin{equation}
K_{ARW}(A, \Omega) = \frac{6A}{\Omega}\frac{1+8e^{A/\Omega}+e^{2A/\Omega}}{-1+e^{2A/\Omega}}
\label{eqKarw}
\end{equation}
is the kurtosis of an ARW. This lower bound also has $\Omega-1$ kinks at points $A_i^K$, which are the solution of the equation $K_{ARW}(A_i^K,i)=K_{ARW}(A_i^K,i+1)$. For $A_{i-1}^K\le A\le A_{i}^K$ the lower bound is given by $K_{ARW}(A, i)$, where $i=1,\ldots,\Omega-1$ and for $i=1$ we have $A_{i-1}^K=0$. In each of the $\Omega$ pieces of the lower bound the Kurtosis reaches an absolute minimum independent of $A$ and $\Omega$, which is given by $K_{min}\approx 25.0913$. These features of the lower bound are illustrated in Fig. \ref{fig4}(a) for the case $\Omega=4$.

The upper bound on Kurtosis is shown in Fig. \ref{fig4}(b). This upper bound has a kink at a point $A^K_*$ that depends on system size $\Omega$. The values of $A^K_*$ up to $\Omega=8$ are given in Table \ref{tab1}. 
For $0\le A\le A^K_*$ the upper bound is given by $S_{ARW}(A,\Omega)$. For $A> A^K_*$ we can only determine the upper bound numerically. The asymptotic form of the upper bound for large affinity $A$ is $6A$, which corresponds to
the asymptotic form of $K_{ARW}(A,1)$ in Eq. \eqref{eqKarw}. 

In the linear response regime the lower and upper bounds for kurtosis also tend to the same value, as shown in Fig. \ref{fig1}. In this regime the kurtosis has a fixed value given by $K=30$. This value can be obtained by taking the limit $A\to 0$ of
 $K_{ARW}(A, \Omega)$ in Eq. \eqref{eqKarw}.

 \subsection{Inference of network topology}
 
The idea of statistical kinetics \cite{moff14} is to infer the topology of a network of states in an enzymatic reaction from data gathered in single molecule experiments. While we focus on the entropy production, which has  increments  
$\theta_{\Omega,1}=-\theta_{1,\Omega}=A$, the current analyzed in statistical kinetics has increments $\theta_{\Omega,1}=-\theta_{1,\Omega}=1$. We define the randomness parameter, skewness and kurtosis associated with this current with increments $1$ (instead of $A$) as $R^*$, $S^*$ and $K^*$, respectively. They are related to the same quantities for the entropy production in the following way $R= A R^*$, $S= A^{1/2} S^*$, and $K= A K^*$. 

A main bound in statistical kinetics is $R^*\ge 2/\Omega$, which is valid in the limit $A\to \infty$. One can infer the unknown number of states in an enzymatic reaction by measuring $R$ and calculating the lower bound on the number of states $2/R$. As shown in \cite{bara15d}, there are also lower bounds on skewness and kurtosis in the limit $A\to \infty$, which are $S^*\ge 2/\Omega^{1/2}$ and $K^*\ge 6/\Omega$. These bounds are consistent with the more general lower bounds in Eq. \eqref{eqsmin} and Eq. \eqref{eqkmin}. It turns out that the randomness parameter is the most effective for a direct estimation of the minimal number of states \cite{bara15d}.

These bounds hold in the limit $A\to \infty$. The lower bounds we found here for $S$ and $K$  are more general and take the affinity $A$ into account. Hence, they can be potentially more effective to infer the number of states. The investigation on how to use the bounds conjectured here  for skewness and kurtosis to infer the number of states of a unicyclic network is beyond the scope of this paper. However, we can make the following statements. 
For $A\le A_1^S$ for skewness and for $A\le A_1^K$ for kurtosis the lower bound is dominated by an ARW with $\Omega=1$, hence the lower bounds cannot be used to infer number of states in this regime. The upper bounds for both skewness and kurtosis, for affinities smaller than the affinity for which the kink in the upper bounds take place, can, in principle, be used to infer the number of states. However, differences in the upper bounds  due to the number of states are quite small, as shown in Fig. \ref{fig2}(b) for $S$ and Fig. \ref{fig4}(b) for $K$. Therefore, using the upper bound to infer the number of states does not look promising.    

\section{Violation of bounds in a multicyclic network}
%%%%%%%%%%%%%%%%%%%%%%%%%%%%%%%%%%%%%%%%%%%%%%%%%%%%%%%%%%%%%%%%%%%%%%%%%%%%%%%%%%%%%%%%%%%%%%%%%%%%%%%%%%%%%%%%%%%%%%%%%%%%%%%%%%%%%%%%%%%%%%%%%%%%%%%%%%%%%%%%%%%%%%%%%%%%%%%%%%%%%%%%%%%%%%%%%%%%%%%%%%
 \label{sec5}
 \begin{figure}
\centering\includegraphics[width=75mm]{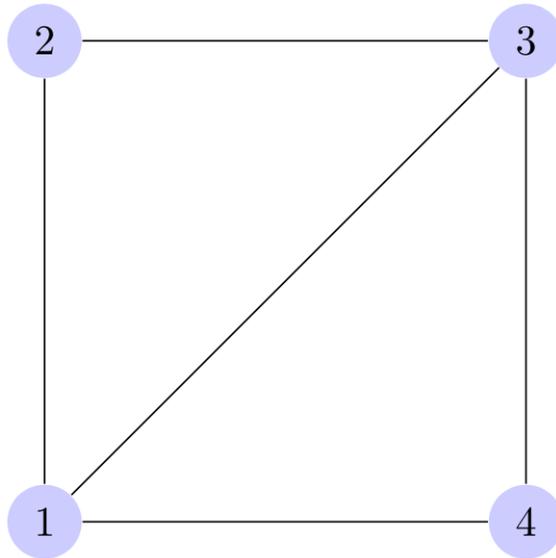}
\vspace{-3mm}
\caption{Multicyclic Network with 4 states. The links between states represent non-zero transition rates.}
\label{fig5}
\end{figure} 

 We now show that the lower bounds we found for unicyclic networks do not hold for multicyclic networks. We consider the multicyclic network with 4 states shown in Fig. \ref{fig5}. We have calculated skewness and kurtosis  for the entropy production  using the methods explained in Sec. \ref{sec3}. The increments for the entropy production for general multicyclic networks are given by $\theta_{ij}=\ln(k_{ij}/k_{ji})$.  We consider the skewness and kurtosis associated with the first passage time distribution of the entropy production.
 
 Defining an affinity dependent bound for multicyclic networks is not so straightforward since there is more than one affinity for multicyclic networks. However, we can check whether for a multicyclic network  $S$ and $K$ can cross the absolute minima, independent of  affinity and system size, for  unicyclic networks, which are given by $S_{min}\approx 3.90973$ and $K_{min}\approx 25.0913$.  
 
 \begin{figure}
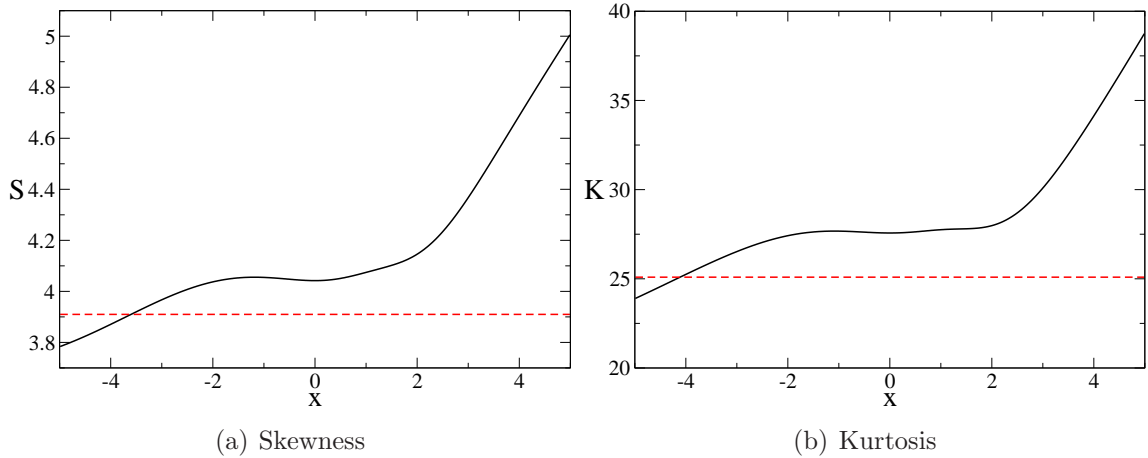

\subfigure[Skewness]{\includegraphics[width=75mm]{Fig6a.eps}}
\subfigure[Kurtosis]{\includegraphics[width=75mm]{Fig6b.eps}}
\vspace{-3mm}
\caption{Skewness $S$ and kurtosis $K$ plotted as  functions of the parameter $x$ for the model with a multicyclic network. The transition rates are parametrized as $k_{12} = k_{23} = k_{34} =k_{41} = \e^{5/4}$,
$k_{21}= k_{13} = k_{32} = k_{43} =  k_{14} = 1$, and $k_{31} = \e^x$. The horizontal dashed lines mark the absolute minimum values $S_{min}\approx 3.90973$ and $K_{min}\approx 25.0913$.}
\label{fig6}
 \end{figure} 

As shown in Fig. \ref{fig6},  they do cross these absolute minima for the multicylic network considered here. Hence, the lower bound for unicyclic networks  does apply to multicylic networks. For the results shown 
in Fig. \ref{fig6}, there is a region for which $S< S_{min}$ but $K>K_{min}$. However, in our numerical investigation  we also found regions for which  $K<K_{min}$ and $S>S_{min}$. The crossing of at least one of the two absolute 
minima constitute a rather generic toll to infer whether a network of states is not unicyclic, which is also a relevant problem in statistical kinetics \cite{moff14}.

%\begin{figure}
%\subfigure[]{\includegraphics[width=91mm]{Fig1a.pdf}\label{fig1a}}
%\subfigure[]{\includegraphics[width=75mm]{Fig1b.pdf}\label{fig1b}}
%\vspace{-3mm}
%\caption{Brownian clock with three states. (a) Particular transition rates for the calculations. (b) Contour plot of $D_\tau/D_J$ for $\gamma_f=1$, where $\gamma_s=\textrm{e}^x$.
%The black lines represent the values $D_\tau/D_J= 0.25, 0.5, 1.0, 1.5$. 
%}
%\label{fig1}
%\end{figure}

%%%%%%%%%%%%%%%%%%%%%%%%%%%%%%%%%%%%%%%%%%%%%%%%%%%%%%%%%%%%%%%%%%%%%%%%%%%%%%%%%%%%%%%%%%%%%%%%%%%%%%%%%%%%%%%%%%%%%%%%%%%%%%%%%%%%%%%%%%%%%%%%%%%%%%%%%%%%%%%%%%%%%%%%%%%%%%%%%%%%%%%%%%%%%%%%%%%%%%%%%%
\section{Conclusion}
%%%%%%%%%%%%%%%%%%%%%%%%%%%%%%%%%%%%%%%%%%%%%%%%%%%%%%%%%%%%%%%%%%%%%%%%%%%%%%%%%%%%%%%%%%%%%%%%%%%%%%%%%%%%%%%%%%%%%%%%%%%%%%%%%%%%%%%%%%%%%%%%%%%%%%%%%%%%%%%%%%%%%%%%%%%%%%%%%%%%%%%%%%%%%%%%%%%%%%%%%%
\label{sec6}

We introduced a method to calculate the cumulants associated with the first passage time distribution of an arbitrary current, or more generally associated with any observable of the form in Eq. \eqref{eqgencurr}. 
Our method circumvents the problem of evaluating the full scaled cumulant generating function $\rho(s)$, which can only be obtained in terms of the transition rates  for quite simple models. Instead the cumulants are
obtained in terms of certain coefficients that are much easier to evaluate. 

The skewness and kurtosis related to the first passage time distribution of entropy production for unicyclic networks have been analyzed with our method. We conjectured lower and upper bounds on these standardized moments. Interestingly, the lower bounds have several kinks, which comes from  the fact that an ARW with $\Omega$ states does not minimize the skewness and kurtosis for any value of the  affinity. This lower bound is different from a previously known lower bound associated with the second cumulant 
\cite{bara15}, which has no kinks and is minimized  for an ARW with $\Omega$ states.
  
Skewness and kurtosis for unicyclic networks have absolute minima, independent of affinity and number of states. We have shown that for a multicylic network both skewness and kurtosis can go below the minima for unicylic networks. 
Hence, multicylic networks are not bounded by the bounds we conjectured for unicylic networks. Crossing of these minima provides a generic tool to infer whether an underlying network of states is indeed multicylic.

As an interesting perspective for future work, the application of our bounds to statistical kinetics could lead to new ways to obtain information about an enzymatic scheme  from data obtained in single molecule experiments. Our bounds 
take thermodynamic affinity into account, which, in principle, can be controlled in an experiment.

%%%%%%%%%%%%%%%%%%%%%%%%%%%%%%%%%%%%%%%%%%%%%%%%%%%%%%%%%%%%%%%%%%%%%%%%%%%%%%%%%%%%%%%%%%%%%%%%%%%%%%%%%%%%%%%%%%%%%%%%%%%%%%%%%%%%%%%%%%%%%%%%%%%%%%%%%%%%%%%%%%%%%%%%%%%%%%%%%%%%%%%%%%%%%%%%%%%%%%%%%%
\section*{References}

\bibliographystyle{iopart-num}

\bibliography{refs}  

%==========================================================================
%{\noindent \textbf{Acknowledgements}}\newline We thank 
%==========================================================================

\end{document}